\documentclass[conference]{IEEEtran}
\IEEEoverridecommandlockouts
\usepackage[noadjust]{cite}
\usepackage{amsmath,amssymb,amsfonts}
\usepackage{algorithmic}
\usepackage{graphicx}
\usepackage{textcomp}
\usepackage{xcolor}
\usepackage{hyperref}
\usepackage{subcaption}
\usepackage{tikz}
\usepackage[american]{circuitikz}

\usetikzlibrary{shapes,arrows,positioning,calc}

\def\BibTeX{{\rm B\kern-.05em{\sc i\kern-.025em b}\kern-.08em
    T\kern-.1667em\lower.7ex\hbox{E}\kern-.125emX}}
\begin{document}

\tikzset{
block/.style = {draw, fill=white, rectangle, minimum height=3em, minimum width=3em},
tmp/.style  = {coordinate}, 
sum/.style= {draw, fill=white, circle, node distance=1cm},
input/.style = {coordinate},
output/.style= {coordinate},
pinstyle/.style = {pin edge={to-,thin,black}
}
}

\title{Non-Linear Modeling and Analysis of Amplifier-Less Potentiostat Architectures\\
\thanks{This work has been submitted to the IEEE for possible publication. Copyright may be transferred without notice, after which this version may no longer be accessible.}
}

\author{\IEEEauthorblockN{Andrea Sannino\textsuperscript{1,*}, David-Peter Wiens\textsuperscript{2}, Maurits Ortmanns\textsuperscript{2}, José I. Artigas\textsuperscript{1}, Aránzazu Otín\textsuperscript{1}}
\IEEEauthorblockA{\textsuperscript{1}Group of Power Electronics and Microelectronics, I3A Institute, University of Zaragoza, Spain \\
\textsuperscript{2}Institute of Microelectronics, University of Ulm, Germany\\
\textsuperscript{*}Correspondence: asannino@unizar.es}
}

\maketitle

\begin{abstract}
In this article, a previously published amplifier-less potentiostat architecture is further examined. Starting with a linearized model, the impact of the most important parameters is studied taking in account the electrodes-solution electrochemical interface. A detailed model is obtained and thoroughly verified, and recommended operating conditions are given for certain limit load conditions. Then, a more complete non-linear model is developed to take in account the measurement uncertainty introduced by the circuit non-linear components. This non-linear model is compared to a time domain description of the circuit and it is verified that it can predict the non-linear behavior with a precision better than 20\%. This result enables the circuit designers to compensate for these effects and ultimately reduce the overall measurement uncertainty.
\end{abstract}


\section{Introduction}
\begin{figure}
    \centering
    \includegraphics{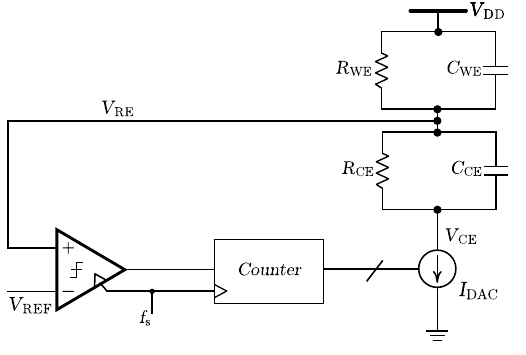}
    \caption{Circuit schematic of the potentiostat proposed in \cite{akram_36_2024,akram_ultra-low-power_2025}.}
    \label{fig:circ_schem}
\end{figure}

Biosensors have become increasingly popular in recent years for studying and measuring electrochemical reactions. In particular, enzyme-based biosensors are the most commonly used for glucose and lactate detection \cite{kieninger_microsensor_2018}.These devices consist of electrodes with enzymes immobilized directly on their surface that can produce an electrical signal related to the target substance concentration. High miniaturization capability, CMOS compatibility, and cost make this technology one of the most attractive in many biosensing fields. Chronoamperometry (\textit{CA}) is one of the most popular techniques to interrogate these biosensors \cite{ying_current_2021}. It consists of three electrodes submerged in a medium in which the target substance is added. A fixed voltage difference is applied between the working electrode (\textit{WE}), where the enzyme is immobilized, and the reference electrode (\textit{RE}), where no current is drawn thanks to the electrode material and dimensions \cite{merrill_electrical_2005}. The current flowing (\textit{I}\textsubscript{SENSE}) between the \textit{WE} and counter electrode (\textit{CE}) is monitored. When the substance concentration changes, \textit{I}\textsubscript{SENSE} changes accordingly. The electrodes are modeled as an \textit{RC} network where the resistance models the so-called Faradaic effects, where actual charge transfer happens at the electrode interface, and the capacitor models processes where only charge redistribution effects occur, and no charges actually pass \cite{ying_current_2021, merrill_electrical_2005}. Potentiostats were introduced to carry out this measurement, with the need to measure currents in the nA-$\mu$A range \cite{hsu_current-measurement_2018}. Many potentiostat topologies have been developed in recent years, converting the sensing current into various domains such as voltage, using trans-impedance amplifiers (\textit{TIAs}) \cite{hsu_current-measurement_2018, lin_212_2023, kim_12_2016, massicotte_cmos_2016, tang_integrated_2020, nazari_implantable_2014}, or frequency \cite{lu_19_2021,yu_reconfigurable_2023,lu_184_2021,li_1_2023,son_low-power_2017}. In \cite{akram_36_2024,akram_ultra-low-power_2025} a digital low-dropout regulator (\textit{DLDO}) like potentiostat was proposed for the first time, comprising only a comparator, a current digital-to-analog converter (\textit{DAC}), and a digital controller. The system architecture is shown in Fig.\ref{fig:circ_schem} and consists of a digital control loop that monitors the voltage at the \textit{RE} and compares it to a reference value with a discrete-time comparator. The output of the comparator controls an up-down counter whose digital value fixes the output current of a current \textit{DAC}. This current is then fed to the electrodes to sustain the electrochemical reaction. 
The load at the \textit{CE} is responsible for the \textit{V}\textsubscript{CE} waveform, while the \textit{WE} load is responsible for the controlled \textit{V}\textsubscript{RE} waveform. For \textit{CA} it is of key importance to maintain a controlled voltage difference between the \textit{WE} and the \textit{RE}, hence only this quantity is monitored and controlled by the loop, and the voltage at the \textit{CE} is considered a consequence of the regulation. 

Although the system was proven to be suitable for the application, a more detailed stability analysis needs to be carried out to study the parameters affecting stability and how designers can alleviate these problems. Furthermore, it is of key importance to consider the system interaction with the electrodes and medium and how these factors impact performance. In Section \ref{secStabAnaLin} a detailed linear model is found with the assumptions made in \cite{akram_36_2024,akram_ultra-low-power_2025}. In Section \ref{secModValid} the model is verified and the recommended operating conditions are given. Then, in Section \ref{secNonLinStabAna} the time response is analyzed and the model is further complicated adding non-linear effects and estimating their behavior with the describing function method. This enables an improvement in the digital control of the potentiostat to reduce the drawbacks coming from the non-linearities, improving the system's performance.
\section{Stability Analysis} \label{secStabAnaLin}
\subsection{System Transfer Function Calculation} \label{secTransFunct}
\begin{figure}
    \includegraphics{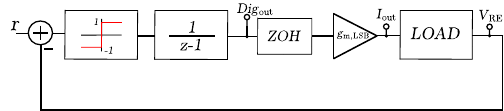}
    \caption{Block Diagram of the system. The reference is fed to the comparator and the counter, $V_{\mathrm{RE}}$ is established as the DAC output times the load.}
    \label{fig:Block_diagram}
\end{figure}

Fig.\ref{fig:Block_diagram} shows the block diagram of the system. To find an approximate transfer function, the discrete-time comparator is modeled as a block with gain of 1 (as done in \cite{akram_36_2024,akram_ultra-low-power_2025}), the counter as a discrete-time (DT) integrator, and the DAC as a zero-order hold (ZOH) followed by an equivalent gain of $g_\mathrm{m,LSB}$. The DAC is assumed with a LSB of $I_\mathrm{LSB}$. When the output of the digital counter is normalized to an LSB voltage of $1$ V, the combination can be expressed as an LSB transconductance of $g_\mathrm{m,LSB}=I_\mathrm{LSB}/1V$. The DAC and the continuous-time load were discretized using the ZOH method \cite{franklin_feedback_2015}. This leads to the following load transfer function:
\begin{equation}
    Z_{LOAD} (z) = \frac{R_{\mathrm{WE}}(1-e^{-\tfrac{T_\mathrm{s}}{\tau}})}{z-e^{-\tfrac{T_\mathrm{s}}{\tau}}} \label{eqZload}
\end{equation}
where $T_\mathrm{s}$ is the sampling period, and $\tau = R_{\mathrm{WE}}C_{\mathrm{WE}}$. Calculating the open-loop transfer function, by simply multiplying this quantity by the digital filter transfer function times $g_\mathrm{m,LSB}$, we obtain the following open-loop response:
\begin{equation}
    G_{OL} (z) = \frac{g_\mathrm{m,LSB}R_{\mathrm{WE}}(1-e^{-\tfrac{T_\mathrm{s}}{\tau}})}{(z-1)(z-e^{-\tfrac{T_\mathrm{s}}{\tau}})} \label{eqOL}
\end{equation}
Equation \eqref{eqOL} shows the presence of two poles: one in $z = 1$, and another placed between $0<z<1$ due to the load transfer function.
\subsection{Effect of system parameters on stability}\label{secParamStab}
Since the \textit{DAC} equivalent transconductance and the sampling frequency are design parameters, it is interesting to study their impact on the system stability. The effect of the  design parameters over the closed-loop poles can be studied by separating the open-loop transfer function into a variable gain $K=g_\mathrm{m,LSB}R_{\mathrm{WE}}(1-e^\frac{-T_\mathrm{s}}{\tau})$ and the following transfer function:
\begin{equation}
    L(z) = \frac{1}{(z-1)(z-e^{-\tfrac{T_\mathrm{s}}{\tau}})} \label{eqLoopGain}
\end{equation}
For $K \approx 0$ the system has two roots on the positive real axis of the z-plane, one in $z = 1$ due to the digital counter and another caused by the load on the working electrode.
Increasing $K$, the two roots move towards each other on the real axis up to a value where they join. From there, any gain increase will lead to two complex conjugate poles whose imaginary part is directly related to the gain. Regarding system stability, it is interesting to find the gain value for which the two roots lay on the unit circle, since this is the system stability limit. Imposing $|z_{roots}|=1$, it is possible to find that the stability limit is: 
\begin{equation}
    K \le K_1 = 1-e^{-\tfrac{T_\mathrm{s}}{\tau}} \label{eqK1}
\end{equation}
that ultimately results in:
\begin{equation}
    g_\mathrm{m,LSB}R_{\mathrm{WE}} \le 1 \label{eqstability}
\end{equation}
Any value of $g_\mathrm{m,LSB}$ or $R_{\mathrm{WE}}$ that causes a $K>K_1$ will lead to system instability. $R_{\mathrm{WE}}$ is an electrode parameter that depends on the current flowing through the electrode-solution interface during a specific reaction. It is simply a consequence of the current change when the target substance concentration varies, hence, the circuit designer has no control over it. On the other hand, two key system parameters can be controlled: $f_\mathrm{s}$ and $g_\mathrm{m,LSB}$.

The DAC LSB value acts as a loop gain; therefore, it is important to keep it as low as possible to improve stability. This also helps with resolution, allowing measurement of currents in the pA-nA range. Conversely, it is expected that increasing the sampling frequency would actually worsen system stability. In fact, increasing $f_\mathrm{s}$ will move the dominant pole close to the integrator pole, reducing the phase margin (\textit{PM}) and degrading the system's time response.

Furthermore, the load parameters must be taken into consideration, as they play an important role in system stability. An increase in $\tau$ has the same effect as an increase in the sampling frequency. This indicates that the system exhibits lower PM for large load resistances, $R_\mathrm{WE}$, which is computed dividing the reference voltage and the assumed reaction current. Moreover, increasing $C_{\mathrm{WE}}$ also decreases the stability margins as it is responsible for the second pole's location. This can provide an indication on the types of electrodes suitable for the system. 

\section{Model Validation and Results} \label{secModValid}
The system in Fig.\ref{fig:Block_diagram} has been modeled both in the frequency and in the time domain. In the time domain, the system has been described through high-level behavioral modeling using Verilog-A, while the frequency model has been developed with Matlab transfer functions. The stability theory is verified using these two different representations.
\subsection{Frequency Domain Analysis}
\begin{figure}[htbp]
    \subfloat[]{\includegraphics[]{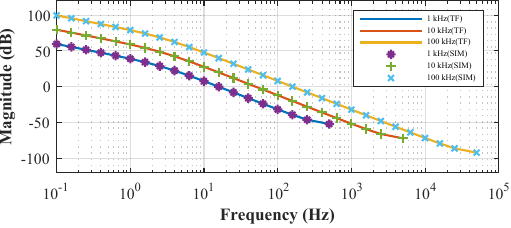}\label{fig:Bode_mag_comp_figa}} \\
    \subfloat[]{\label{fig:Bode_phase_comp_figb}\includegraphics[]{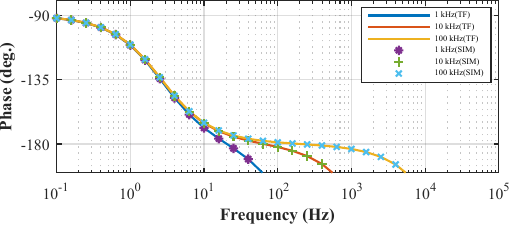}}
    \caption{Bode plots of the calculated transfer functions. Matlab Transfer Function (TF) and Verilog-A (SIM) for $g_\mathrm{m,LSB}=10$ nS (i.e. $I_\mathrm{LSB}=10$ nA), $R_{\mathrm{WE}}=60$ M$\Omega$, $C_{\mathrm{WE}}=1$ nF, $V_{\mathrm{REF}}=0.6$ V and changing $f_\mathrm{s}$. (a) Represents the magnitude while (b) the phase of the Bode plot. }
    \label{fig:Bode_phase_comp}
\end{figure}

Bode plots were generated for the linearized models to study the frequency response. The open-loop transfer function is simulated and analyzed by varying the sampling frequency. The frequency response, shown in Fig. \ref{fig:Bode_phase_comp}, is in accordance with the stability analysis. As predicted, Fig. \ref{fig:Bode_mag_comp_figa} shows a pole at the s-plane origin and a second pole due to the $RC$ load, which changes the gain slope. In this specific case, the dominant pole location is found at a frequency of $\sim3$ Hz and does not depend on $f_\mathrm{s}$. The sampling frequency acts as a loop gain because the continuous-time equivalent of the integrator-accumulator can be modeled by the transfer function $f_\mathrm{s}/s$.
Thus, it can be concluded that as the sampling frequency increases, the system's overshoot will increase due to a decreased PM. Whenever the system measures higher currents, its PM will improve since the second pole will shift to higher frequencies due to a reduction in load resistance. 
Moreover, it is important to note that the case depicted in Fig. \ref{fig:Bode_phase_comp} represents the worst-case scenario for the system's phase margin since the \textit{DAC} measures the minimum possible current, corresponding to a maximum $R_\mathrm{WE}$. In particular, the phase margins are calculated for $1$, $10$, and $100$ kHz, yielding $3.82$, $1.21$, and $0.38$ degrees, respectively. This confirms the point made at the end of section \ref{secParamStab}.

\subsection{Recommended Operating Conditions}
Following these results, the system's damping was shown to be insufficient to meet the measurement requirements for such small currents. A phase margin of at least $20-30$ deg. should be achieved at any current to obtain acceptable time responses. Considering this, this section provides the recommended operating conditions for measuring small currents. The \textit{DAC} is assumed to be 10-bits, with a $I_\mathrm{LSB}=10$ nA (i.e. $g_\mathrm{m,LSB}= 10$ nS). Various load conditions are explored, and the sampling frequency range to achieve a certain phase margin is calculated using the frequency domain model. Table \ref{tab_pm} summarizes the recommended operating conditions for acceptable stability under different load conditions. $R_\mathrm{WE}$ and $C_{\mathrm{WE}}$ are varied to account for different reaction currents and electrode dimensions. The results clearly show that when  $\tau$ is increased by a factor of 10, the sampling frequency must be reduced by the same factor to achieve the same performance. Of particular interest is the case of maximum resistance, where very low sampling frequencies must be used to achieve acceptable margins.

\begin{table}[h]
\caption{Recommended Operating Conditions: 10-bits $I_\mathrm{LSB} = 10$ nA, $V_{\mathrm{DD}}=1.2$ V, and $V_\mathrm{REF}=0.6$ V.}
\begin{center}
\resizebox{0.49\textwidth}{!}{
\begin{tabular}{|c|c|c|c|c|}
\hline
\textbf{Meas.}&\multicolumn{4}{|c|}{\textbf{System Parameters}} \\
\cline{2-5} 
\textbf{Current [nA]} & \textbf{$R_{\mathrm{WE}}$ [M$\Omega$]}& \textbf{\textbf{$C_{\mathrm{WE}}$ [nF]}}& \textbf{$f_\mathrm{s}$ [Hz]}& \textbf{\textbf{PM} [deg.]} \\
\hline
$10$& $60$& $0.1$& $150-350$& $29.2-20.4$ \\
\hline
$10$& $60$& $1$& $15-35$& $29.2-20.4$ \\
\hline
$10$& $60$& $10$& $1.5-3.5$& $29.2-20.4$ \\
\hline
$50$& $12$& $0.1$& $17-40$ k& $31.3-20.8$ \\
\hline
$50$& $12$& $1$& $1.7-4$ k& $31.3-20.8$ \\
\hline
$50$& $12$& $10$& $170-400$& $31.3-20.8$ \\
\hline
\end{tabular}}
\label{tab_pm}
\end{center}
\end{table}

\section{Non-Linearity Analysis and Prediction} \label{secNonLinStabAna}
\subsection{Time Response Analysis} \label{sec:TDAnalysis}
\begin{figure}[h]
    \subfloat[]{\includegraphics[]{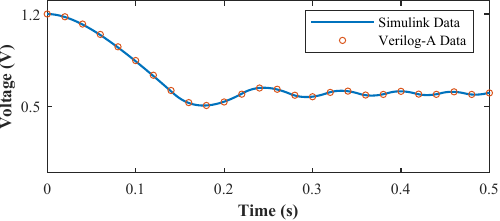}\label{fig:verilog_a_simulink_figa}} \\
    \subfloat[]{\label{fig:verilog_a_simulink_figb}\includegraphics[]{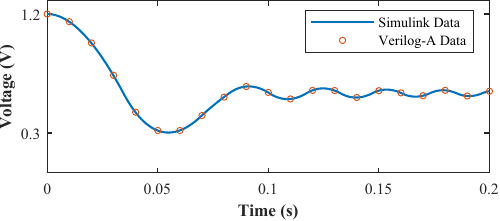}}
    \caption{Comparison between Simulink and Verilog-A non-linear models for $g_\mathrm{m,LSB} = 125$ pS, $R_{\mathrm{WE}}=50$ M$\Omega$, $C_{\mathrm{WE}}=1$ nF, $V_\mathrm{DD} = 1.2$ V and (a) $f_\mathrm{s}=1$ kHz, (b) $f_\mathrm{s}=10$ kHz.}
    \label{fig:Vre_verilogA_simulink}
\end{figure}

Fig. \ref{fig:Vre_verilogA_simulink} shows the results of the time evolution of $V_{\mathrm{RE}}$ obtained with the two models and for two sampling frequencies, adding the comparator non-linearity. While it is true that the system has a faster settling time at a higher sampling frequency, the overshoot also increases significantly. This is predicted by the stability analysis since increasing the sampling frequency results in having a pole closer to the origin, causing a lower damping ratio ($\zeta$).
As a consequence of the lower $\zeta$, we also expect the PM to reduce accordingly. Furthermore, in steady-state, the time response does not resemble that of a linear double pole system. In fact, the output voltage shows periodic oscillations at a fixed frequency and amplitude known as limit cycle oscillations \cite{franklin_feedback_2015}. Since the measurement is taken at the counter output, this limit cycle directly represents a measurement uncertainty and its reduction can be beneficial to ensure a more controlled electrode voltage and cheaper post-processing steps.

\subsection{Non-Linear Potentiostat Modeling} \label{secNonLinMod}
As seen in Section \ref{sec:TDAnalysis}, non-linearities must be taken into account in order to obtain a detailed model of the analyzed potentiostat. In particular, the system exhibits a memory-less non-linearity (i.e. the comparator). For these types of non-linearities, the behavior can be approximated by the describing function method \cite{kochenburger_frequency_1950}. In practice, the non-linearity is excited with a sinusoid and the output Fourier series is studied. By approximating the series to the first harmonic, it is possible to find an approximate equivalent gain for the memory-less non-linearity \cite{truxal1955automatic}. In the case of a comparator, the equivalent gain is \cite{franklin_feedback_2015}:
\begin{equation}
    K_{\mathrm{eq}}(a)=\frac{4N}{\pi a} \label{eqKEQ}
\end{equation}
where $N=1$ in our case and $a$ is the input sinusoid amplitude. By adding this gain and studying the root-locus as a function of $a$ it is possible to predict the amplitude and frequency of the limit cycle. Specifically, finding $a$ and $\omega_\mathrm{{LimCyc}}$ for which the roots lie on the unit circle will provide an approximate value for the amplitude and frequency of the oscillations. Using \eqref{eqOL} and \eqref{eqK1} and recognizing that the roots of the system are at:
\begin{equation}
    r_{1,2}(K')=\frac{(e^{\frac{-T_\mathrm{s}}{\tau}}+1)}{2} \pm j\frac{\sqrt{4(e^{\frac{-T_\mathrm{s}}{\tau}}+K')-(e^{\frac{-T_\mathrm{s}}{\tau}}+1)^2}}{2} \label{eqRoots}
\end{equation}
with $K'=g_\mathrm{m,LSB}R_{\mathrm{WE}}K_{\mathrm{eq}}(a)(1-e^{\frac{-T_\mathrm{s}}{\tau}})$, and that the roots are on the unit circle when $K'= K_1 = (1-e^{\frac{-T_\mathrm{s}}{\tau}})$, we retrieve:
\begin{equation}
\begin{cases}
    a = \frac{4g_\mathrm{m,LSB}R_{\mathrm{WE}}}{\pi} \\
    \omega_\mathrm{{LimCyc}} = f_\mathrm{s}\Im(log(|r_{1,2}(K'=K_1)|))
\end{cases}
\label{eqLimitCycle}
\end{equation}
This system of equations gives an approximate value for the amplitude of the oscillation, $a$, and its frequency $\omega_{\mathrm{LimCyc}}$.
\subsection{Non-Linear Model Verification}
\begin{figure}[h]
    \subfloat[]{\includegraphics[]{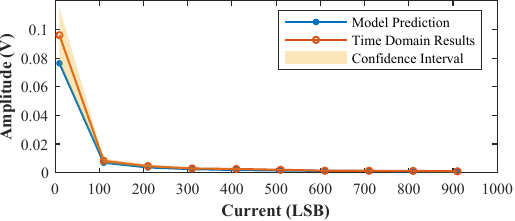}\label{fig:Ampl_LimCyc}} \\
    \subfloat[]{\includegraphics[]{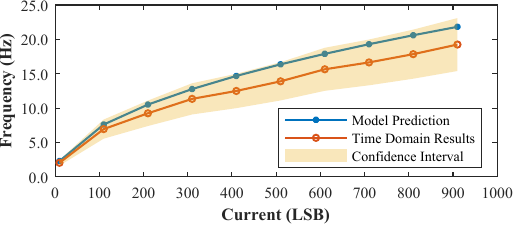}\label{fig:Freq_LimCyc}}
    \caption{Comparison of the non-linear model limit cycle parameters prediction with the time domain model results. (a) Amplitude. (b) Oscillation Frequency.}
    \label{fig:LimCyc}
\end{figure}

To verify the non-linear model found in Section \ref{secNonLinMod}, the time domain model was compared with the transfer functions that included the non-linearity. The \textit{DAC} was assumed to be 10-bit, with an LSB of $125$ pA, $V_{\mathrm{DD}}=1.2$ V, $V_{\mathrm{REF}}=0.6$ V, $C_{\mathrm{WE}}=10$ nF, and $f_\mathrm{s} = 1$ kHz. The reaction current is varied by changing the electrode resistance, and the values of $a$ and $\omega_{\mathrm{LimCyc}}$ are predicted according to \eqref{eqLimitCycle}. Fig. \ref{fig:LimCyc} shows the results of this comparison. This demonstrates that the approximate non-linear model is capable of predicting the limit cycle frequency and amplitude with a precision better than 20\%. With these results, it is possible to design a digital compensator after the comparator to filter out the limit cycle, thus reducing the measurement uncertainty \cite{franklin_feedback_2015}.
\section{Conclusions}
In this work, a recently introduced amplifier-less potentiostat architecture has been thoroughly analyzed, beginning with the topology proposed in \cite{akram_36_2024,akram_ultra-low-power_2025}. First, the stability of the linearized model has been assessed, as proposed in the original work, focusing on investigating all the parameters that affect stability and the time response. Simulations with two different system representations have been carried out to verify the theoretical assumptions that were made. Then, the model has been extended to consider the non-linear effects caused by the comparator. This non-linearity has been proved to be specially relevant for estimating and predicting the key parameters of the limit cycle. According to the obtained results, a more efficient digital control design can be implemented, leading to the limitation of non-linear effects  and reducing measurement uncertainty.  
\bibliographystyle{ieeetr}
\bibliography{MWSCAS_25_bib}
\end{document}